\documentclass[fleqn,usenatbib,useAMS]{mnras}

\usepackage{graphicx}	% Including figure files
\usepackage{amsmath}	% Advanced maths commands
\usepackage{amssymb}	% Extra maths symbols
\usepackage{multicol}        % Multi-column entries in tables
\usepackage{bm}		% Bold maths symbols, including upright Greek
\usepackage{pdflscape}	% Landscape pages
\usepackage{xcolor}

\usepackage[T1]{fontenc}
\usepackage{ae,aecompl}

\usepackage{newtxtext,newtxmath}
\title[A large UV bubble around R Dor]{A large bubble around the AGB star R Dor detected in the UV}

\author[R. Ortiz \& M.A. Guerrero]{R. Ortiz$^{1}$\thanks{Contact e-mail: \href{mailto:rortiz@usp.br}{rortiz@usp.br}} and M.A. Guerrero$^{2}$
\\
$^{1}$Escola de Artes, Ci\^encias e Humanidades, USP, Av. Arlindo Bettio 1000,
03828-000 S\~ao Paulo, Brazil \\
$^{2}$Instituto de Astrof\'\i sica de Andaluc\'\i a (IAA-CSIC),
Glorieta de la Astronom\'\i a s/n, E-18008 Granada, Spain}

\date{Last updated 2020 June 10; in original form 2013 September 5}

\pubyear{2022}

\begin{document}
\label{firstpage}
\pagerange{\pageref{firstpage}--\pageref{lastpage}}
\maketitle

% Abstract of the paper
\begin{abstract}
Many asymptotic giant branch (AGB) and supergiant stars exhibit extended detached shells in the far-infrared, resembling rings or arcs. 
These structures have long been interpreted as the bow shock formed in the interface between the stellar wind and the interstellar medium, the astrosphere. 
To date, only a few AGB stars have been observed showing an extended shell in the ultraviolet: the cometary tail drifting away from $o$ Ceti, and a bubble around IRC+10216, CIT6, and U Hya. This paper describes a search of UV extended shells around AGB stars using archival {\it GALEX} far-UV images. After inspecting visually 282 {\it GALEX} images, we identified the fourth discovery of a UV bubble around the AGB star R Dor. 
The bubble is seen as a $26\arcmin \times 29\arcmin$ ring, corresponding to an actual diameter of $0.41 \times 0.46$ parsec$^2$. The mass of the thin UV bubble is estimated to be $\simeq$0.003 $M_{\odot}$.
The morphological asymmetry (less than $\sim 20$\%) and brightness variations of this shell are uncorrelated with the stellar proper motion and thus they can rather be ascribed to inhomogeneities in the ISM. 
Archival \emph{IRAS} 60 and 100$\mu$m images reveal that the bubble is filled with cold (i.e. $\la 32$K) dust. 
All UV bubbles known to date are limited to be within a distance $\la 350$ pc and at high Galactic latitudes ($|b| \ga 35\degr$), which suggests that their detection is hampered in most cases by the strong UV interstellar extinction.
\end{abstract}

% Select between one and six entries from the list of approved keywords.
% Don't make up new ones.

%%%%%%%%%%%%%%%%%%%%%%%%%%%%%%%%%%%%%%%%%%%%%%%%%%

%%%%%%%%%%%%%%%%% BODY OF PAPER %%%%%%%%%%%%%%%%%%

% The MNRAS class isn't designed to include a table of contents, but for this document one is useful.
% I therefore have to do some kludging to make it work without masses of blank space.
% \begingroup
% \let\clearpage\relax
% \tableofcontents
% \endgroup
% \newpage

%Section 1
\section{Introduction}

The first large scale structure detected in the UV associated with the mass loss of an AGB star was found around $o$ Cet \citep{Martin2007}. Images obtained by the \emph{GALEX} observatory showed an arc-like structure opposed to a long cometary tail extending up to 2$^{\circ}$ from the star \citep{Martin2007}. 
The arc and the tail are both aligned with the direction of the large stellar proper motion, 225.8 mas yr$^{-1}$ \citep[{\it HIPPARCOS},][]{Turon1993}. Actually, the arc seen in $o$ Cet (and detached shells in other AGB stars) can be detected over a wide range of wavelengths, from radio wavelengths to UV, and is formed by the shock between the stellar wind and the local interstellar medium (hereafter ISM) or gas previously expelled from the star \citep{Libert2007,Cox2012}. The infrared radiation emitted by arcs and shells is thermal, after grains are heated by the passage of a shock wave \citep{Cox2012}, whereas the UV emission is probably composed by emission lines of various atomic species. Although the morphology of these structures generally resembles an arc, the fermata symbol or a shell, $o$ Cet is the only case known to date where a {\it drifting or cometary tail} is seen trailing behind the star as it moves at large speed \citep{Knapp2003,Wareing2007}.

Radio and infrared observations of AGB stars have revealed a large number of detached shells (often reaching several arc minutes in size) formed by the shock between the stellar wind and the local ISM \citep{Libert2007,Cox2012,Brunner2019,Mecina2020}. However, after $o$ Cet, only three additional detached shells were eventually detected at UV wavelengths to date: IRC+10216 \citep{SC2010}, CIT\,6 \citep{SM2014}, and 
U\,Hya \citep{Sanchez2015}. Differently from $o$ Cet, all these shells are approximately round, and hereafter they will be called ``bubbles''. In all these cases the AGB star is located near the centre of the expanding bubble, which follows the proper motion of the star.

This paper reports the fourth discovery of a newly found UV bubble around an AGB star. In Sect. \ref{search} we describe the sample of AGB stars examined, and some characteristics of these images; Sect. \ref{mainchar} 
gives a general description of R Dor and its close circumstellar environment; Sect. \ref{results} describes the bubble around R\,Dor, both in the UV and IR; in Sect. \ref{discussion} we discuss the factors that play a role in the {\it detection} of UV shells, as well as those that contribute to their {\it formation and endurance}; in Sect. \ref{conclusions} we present our conclusions.

%Section 2
\section{A search for UV structures around nearby AGB stars}
\label{search}

The {\it GALEX} satellite \citep[{\it Galaxy Evolution Explorer},][]{Morrissey2005} constitutes the main database for this research because it covered large portions of the sky in the near- and far-UV bands. The target sample was composed by: (1) all regular Miras listed in the {\it general catalogue of variable stars} \citep[GCVS5.1,][]{Samus2017}; (2) the list of nearby semiregular stars, compiled by \citet{GvL2007}; (3) the list of AGB and supergiant stars showing bow shocks and detached shells, detected in the far-IR by \citet{Cox2012}. After cross-correlating these lists with the {\it GALEX} database we were left with a total of 282 stars observed by {\it GALEX} in at least one band. Eventually, we inspected visually each image in search of circumstellar features. 
As a result, we have found one previously unknown bubble, associated with the AGB star R Dor.

%Section 2.1
\subsection{Archival UV and IR data}

There is only one {\it GALEX} observation registering R Dor: tilename {\sc ais}$\_ 420$, tilenumber 50420. 
The images were obtained on 28$^{\rm th}$ September 2008 with total exposure times of 206 seconds in the two photometric bands: the far-UV ($1344 - 1786$ \AA) and the near-UV ($1771 - 2831$ \AA). 
The spatial resolution of the images are $4.5\arcsec$ and $6.0\arcsec$, respectively \citep{Morrissey2005}.
The inspection of the far-UV image suggested the presence of diffuse emission from a bubble-like structure around R\,Dor, but at a low signal-to-noise (S/N) ratio.  
To increase the contrast between the bubble emission and the background, the {\it GALEX} images of R\,Dor were adaptively smoothed \citep{2006MNRAS.368...65E} using a circular Gaussian kernel with a size from 1\farcs5 for pixels with S/N$\geq$4 up to a kernel size of 12\arcsec\ for pixels with S/N$\leq$2.  
The maximum kernel size, mostly applied to the background emission, truly reduces its noise, while the minimum kernel size, being smaller than the spatial resolution, preserves the spatial scale of the bubble emission and thus the image quality. The far- and near-UV images are shown in the top panels of Figure~\ref{mosaic}.

Previous observations of bow shocks around red giants and supergiants showed that they are better vizualized in the far-IR \citep{Izumiura1996,Mecina2020}. 
R\,Dor has been observed by several infrared surveys, from near- to far-IR wavelengths, but the near- and mid-IR images (e.g. {\it 2MASS, WISE}) are not very useful because they are dominated by the stellar emission. 
R\,Dor becomes fainter in the far-IR, especially beyond 60 $\mu$m, providing an observational leverage towards the detection of extended emission from cold dust around it.
In the present study, we used \emph{Infrared Astronomical Satellite} (\emph{IRAS}) observations of R Dor obtained at 60 and 100$\mu$m. 
The images, scaled at MJy~sr$^{-1}$, were extracted from the {\it IRAS Sky Survey Atlas} ({\it ISSA}) available at the {\it IRSA} website\footnote{https://irsa.ipac.caltech.edu/Missions/iras.html}. 
Previous determinations showed that the resolution is variable \citep[$3.5\arcmin \sim 4.9\arcmin$ at 60$\mu$m, $4.5\arcmin \sim 6.1\arcmin$ at 100$\mu$m,][]{Wheelock2002}, depending on the orientation of the long axis of the image, and will be re-evaluated in Sect. \ref{infrared}. 
The \emph{IRAS} 60 and 100 $\mu$m images are shown in the bottom panels of Figure~\ref{mosaic}.

%Section 3
\section{Main characteristics of R Dor}
\label{mainchar}

%Section 3.1
\subsection{The stellar component}

R Dor is a M8{\sc iii} semiregular (SRb) pulsating variable, varying its V-band magnitude between 4.8 and 6.6 mag within a period of 338 d \citep{Samus2017}. Apart its normal magnitude range due to its pulsation, the light curve obtained by the {\sc aavso}\footnote{{\it The AAVSO}, www.aavso.org} shows, as observed in some other SR stars, some abnormally intense maxima (4.5 visual mag in 2011 and 2018$-$2019) and minima ($7.2 \sim 7.4$ visual mag between 1949 and 1952). An analysis of its light curve using the {\it wavelets} method showed that the pulsation seems to switch between the first and third overtone modes, corresponding to the periods of 332 d and 175 d \citep{Bedding1998}. According to \citet{KH1992}, after considering the pulsation period and the $JHKLM$ photometry, R Dor is a `red' semiregular, i.e. an AGB star in the thermal pulse phase, showing characteristics (pulsation period, colour indices, vertical distribution and number density in the Galaxy) very similar to those of Mira-type stars.

R Dor is situated at the distance of 54.6 parsecs \citep{vanLeeuwen2007}, which makes it the nearest AGB star to the Sun. Its proximity contributes to its large proper motion: ${\mu}_{\alpha}=-69.4$ mas yr$^{-1}$, ${\mu}_{\delta}=-75.8$ mas yr$^{-1}$. At the distance of 54.6 pc, the combined proper motion of 102.8 mas~yr$^{-1}$ corresponds to a tangential velocity of 26.6 km~s$^{-1}$ towards PA $222\degr$, i.e. approximately in the southwest direction. Its galactic space velocity relative to the local standard of rest is $V_{UVW} =$ 37.2 km s$^{-1}$.

The small distance makes it also its stellar apparent diameter to be the largest ever measured besides the Sun: $58.7 \pm 2.6$ mas \citep[average value, assuming a uniform disc]{Bedding1997}, which corresponds to an actual average radius of 345 R$_{\odot}$. 
Its spectrum corresponds to a stellar temperature between 2230 K \citep{JudgeStencel1991} and 2710 K \citep{Ohnaka2019}, and its mass has been estimated in the range $M_*/M_{\odot}=0.7 \sim 1.5$ \citep{JudgeStencel1991,Ohnaka2019}.

R Dor was formerly assigned as a binary system, with a 11.9 mag secondary of unknown spectral type at 32.3\arcsec from the primary \citep{Proust1981}, but eventually it was reclassified as an optical pair \citep{Mayer2013}. According to \citet{OG2016} AGB stars showing a far-UV counterpart are binaries because, under normal circumstances, the chromospheric emission is too faint to be detected in the far-UV. R Dor does not have a {\it GALEX} far-UV counterpart, which reinforces the absence of a stellar companion. However, there is some evidence that a substellar companion might actually exist. For example, {\it ALMA} high-resolution observations of its extended atmosphere showed that R Dor rotates two orders of magnitude faster that the expected for a single AGB star \citep{Vlemmings2018}. \citet{Homan2018} argue that R Dor has a circumstellar disc formed by the interaction with an object with at least 2.5 earth masses, situated at 6 AU.

%Figure1
\begin{figure*}
\includegraphics[width=0.995\textwidth]{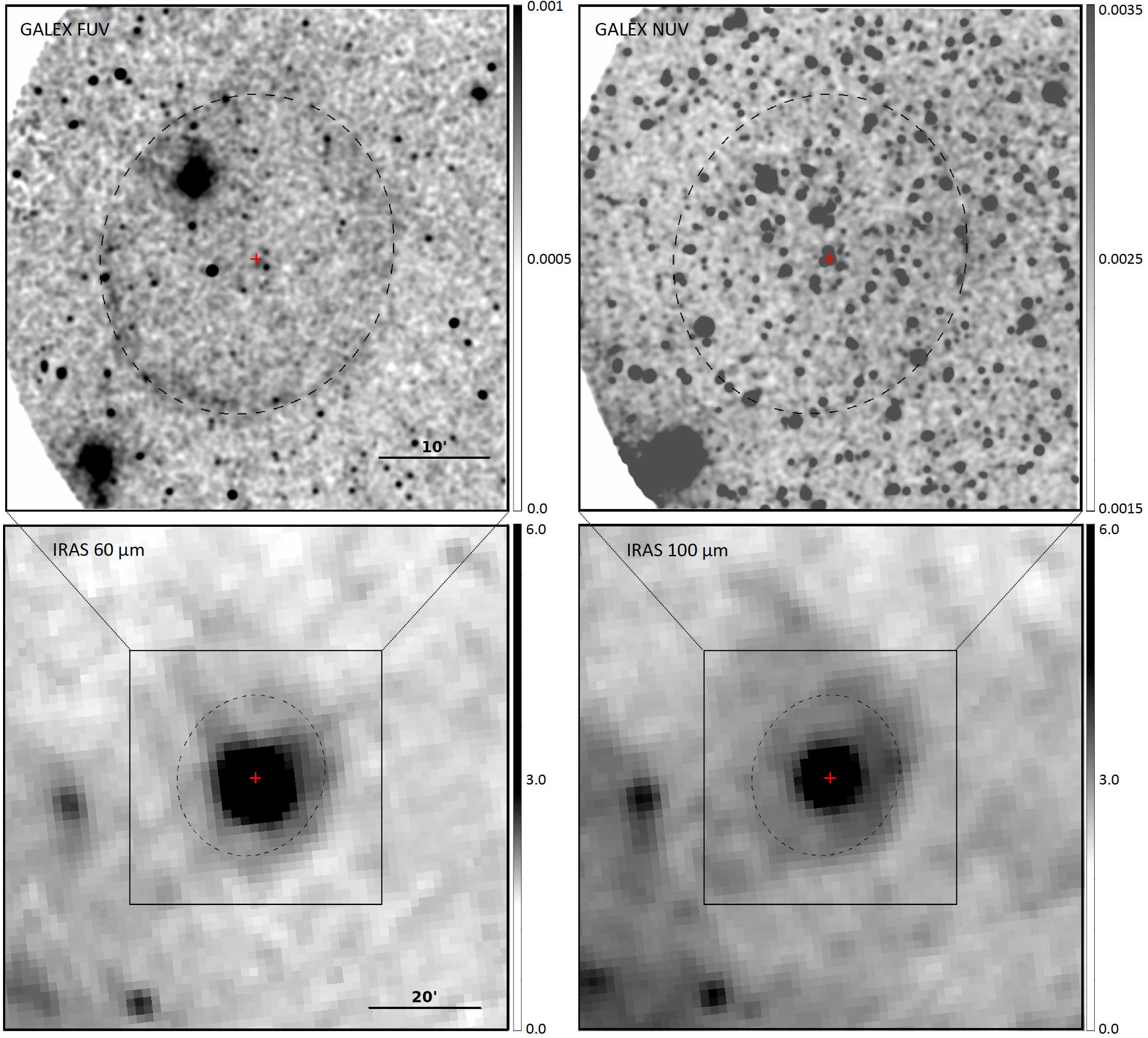}
\caption{Greyscale {\it GALEX} far-UV (top-left) and near-UV (top-right) and {\it IRAS} 60 $\mu$m (bottom-left) and 100 $\mu$m (bottom-right) images of R\,Dor. The red cross marks the position of R\,Dor. Images are shown using a squared-root scale with the side-bars in units of counts~pixel$^{-1}$ for the {\it GALEX} images and MJy~str$^{-1}$ for the {\it IRAS} images. A larger field of view is used for the {\it IRAS} images to display the large-scale variations of the background surface brightness. The field of view of the {\it GALEX} images is highlighted to ease the comparison. The dashed $26'\times29'$ ellipse centred at R\,Dor with major axis along PA = 160$^\circ$ marks the position of the shell. North it top, east to the left.}
\label{mosaic}
\end{figure*}

%Section 3.2
\subsection{Circumstellar matter}
\label{circumstellar}

\citet{Khouri2016} detected the formation of clumpy dust clouds as close as $\sim 1.5$ stellar radii using polarimetric observations. 
Eventually, \citet{Ohnaka2019} reported the presence of a bright spot over the surface of the star, which might be associated with convective cells of circumstellar matter \citep{Vlemmings2019}. 
The mass-loss rate has been estimated to be $(6 - 9) \times 10^{-8}$ M$_{\odot}$~yr$^{-1}$ from visual emission line spectra \citep{Hagen1983} and $(10 - 16) \times 10^{-8}$ M$_{\odot}$~yr$^{-1}$ from CO rotational lines \citep{Lindqvist1992,Olofsson2002,Maercker2016}. 
R Dor exhibits a wide variety of circumstellar emission lines, such as SiO \citep{Balister1977,Olofsson1998}, CO, HCN \citep{Lindqvist1992,Olofsson1998}, SiS and SO \citep{Olofsson1998}, SO$_2$, CN, PN, and PO \citep{BO2018}. The molecular envelope expands at a relatively low velocity: $5.5 - 6.0$ km s$^{-1}$ \citep{Maercker2008,Decin2018}. More recently, the modeling of the SiO emission in its close environment with the {\it ALMA} facility at high-resolution showed evidence of a close circumstellar disc with an outer radius of $\sim 25$ AU and an inclination angle of $110^{\circ}$ \citep{Homan2018}.

%Section 4
\section{Results}
\label{results}

%Section 4.1
\subsection{UV emission}

The top panels of Figure~\ref{mosaic} shows {\it GALEX} UV images of R\,Dor.  
The bubble is conspicuous in the far-UV image, but mostly indistinguishable in the near-UV one, which will no longer be discussed. The location of R\,Dor is marked by a red cross and that of the shell is highlighted by a $26'\times29'$ ellipse whose major axis is oriented along PA $\approx$160$^\circ$ (see below). 

Figure \ref{profile} shows the far-UV surface brightness profile (i.e., count~s$^{-1}$~pix$^{-1}$) along radial rectangular apertures $3\arcmin$-wide from R\,Dor 
% The aperture is positioned with the central star at one of its extremities, and 
oriented at selected azimuthal angles from PA $= 90\degr$ (i.e. eastward) to PA $= 360\degr$ (northward). 
% This plot depicts the brightness profile of the bubble along the radial distance from the star, for various azimuthal angles. 
Unfortunately, the low signal-to-noise ratio and background point sources do not allow us to assure whether the bubble is filled with UV radiation, although this does not seem to be the case as the UV intensity inside and outside the average radius is, considering the uncertainties, similar. 
The thickness of the bubble is variable: it is sharper (only $\sim 2\arcmin$) at PA $= 90\degr$ and broader at PA $=270\degr$ and $360\degr$ ($\sim 4.5\arcmin$). 
The low signal-to-noise ratio hampers a reliable determination of the shape of the profile of the bubble along its the pre- and post-shock zones.

Figure~\ref{radial} shows radial profiles of the bubble along different azimuthal (PA) angle. 
Its average radius is approximately 13\farcm3, with a notable excursion along 
PA $= 360\degr$ (radius of $\sim 16.5\arcmin$), which determines the 14.5$'$ semi-major axis of the ellipse shown in Figure~\ref{mosaic}. The shell thickness also varies notably along different PAs, being sharper towards the South-Southeast and thicker towards the North-Northwest. Figure~\ref{radial} also shows the intensity of the {\it GALEX} far-UV emission, integrated along its thickness. The bubble looks brighter towards PA $= 145\degr$ and $270\degr$, which do not correspond to the direction of the proper motion of the star (PA $= 222\degr$). This case is similar to U Hya, which has a very round UV bubble, without evidence of asymmetry caused by its proper motion \citep{Sanchez2015}. Therefore, in these two cases the azimuthal asymmetries, like those shown in Figs. \ref{profile} and \ref{radial}, are more likely to be due to inhomogeneities in the local ISM, which might affect the expansion of the shell. 

%Section 4.2
\subsection{Infrared emission}
\label{infrared}

\begin{figure}
\includegraphics[width=0.475\textwidth]{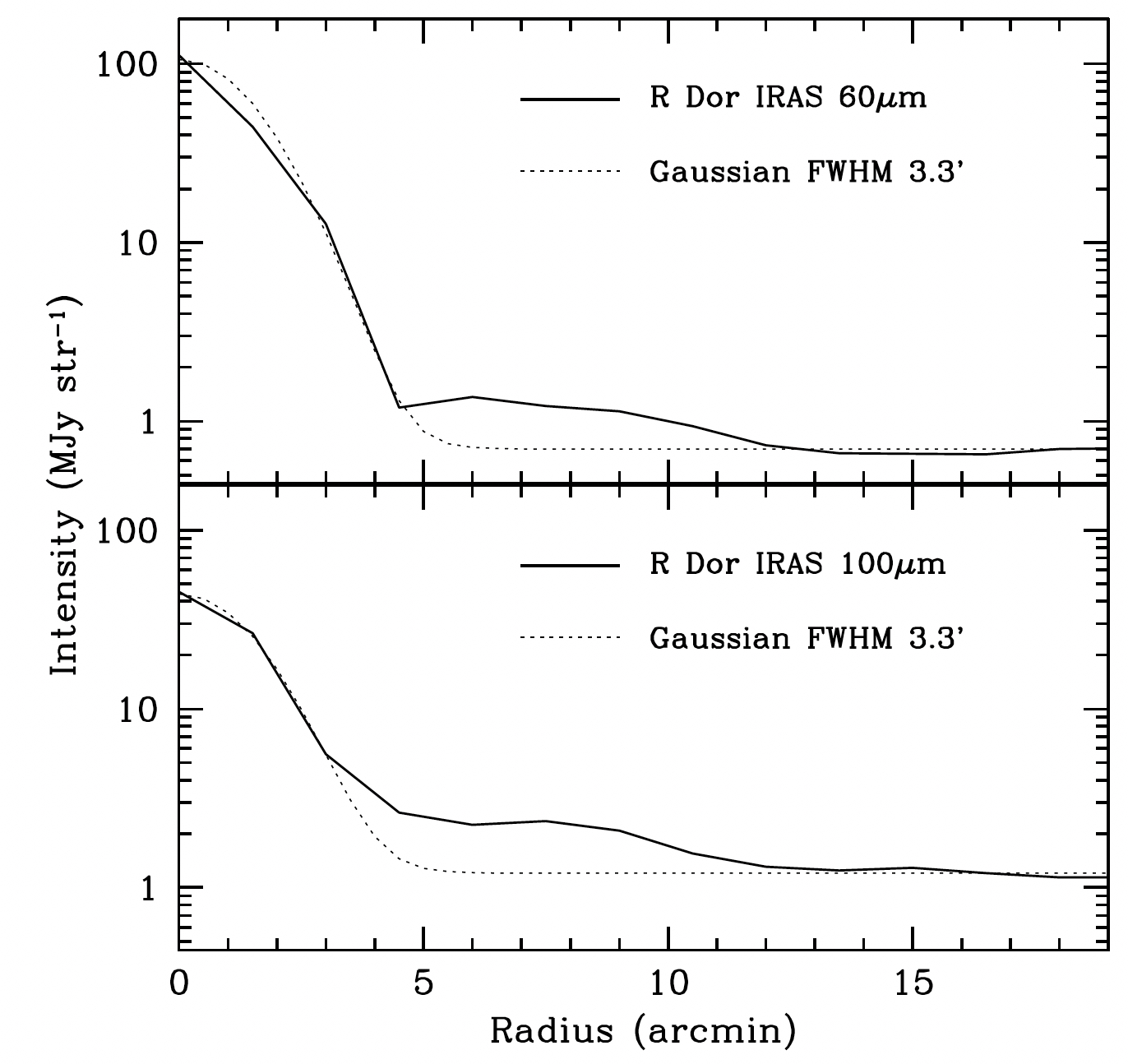}
\caption{
Azimuthally averaged {\it IRAS} 60 $\mu$m (top) and 100 $\mu$m (bottom) spatial profiles of R\,Dor (solid line).  
For comparison, the point spread function of a point source, assumed to have a Gaussian profile normalized to the peak intensity with a FWHM of 3\farcm3, is shown (dotted lines). 
}
\label{iras}
\end{figure}

The far-IR \emph{IRAS} images of R\,Dor seem to show diffuse emission with a similar spatial extent to that of the UV bubble (Fig.~\ref{mosaic}-bottom). 
This is confirmed by the azimuthally averaged profiles at 60 and 100 $\mu$m of IRAS\,04361$-$6210, the infrared counterpart to R Dor, which are consistent with a point source with extended shoulders (Fig. \ref{iras}). As a comparison, an examination of the {\it IRAS} 60 and 100 $\mu$m spatial profiles of two isolated AGB stars, namely EP\,Aqr and SW\,Vir, did not reveal extended shoulders.  
Instead, these can be fitted by a Gaussian profile with a FWHM of 3\farcm3, which is also suitable for the core of R\,Dor. 
This comparison then confirms that there is a circumstellar component extending up to $\sim 13\arcmin$ from the central position, which matches the radius of the UV bubble.  
Using the spatial profiles shown in Figure~\ref{iras}, the flux over the background emission has been integrated in an annular region with inner radius 6\arcmin\ and outer radius 13\arcmin\ to find that the emission in the 100 $\mu$m band (10.2 Jy) is indeed brighter than in the 60 $\mu$m band (6.8 Jy).  
Although the assessment of the inhomogeneous background emission and the contamination of the bright core of the central source make difficult an accurate determination of the {\it IRAS} fluxes from the extended component, the noticeably larger 100 $\mu$m emission suggests that the dust is cold (T$_{\rm dust}\la$32 K).

Like R Dor, \citet{Sanchez2015} observed that the UV and far-IR emission of the bubble around U Hya generally coincide in position. In the case of U Hya, despite the 70 $\mu$m emission is more concentrated along the UV ring, it is also detected in its interior. 
We conclude that the UV bubble around R\,Dor might be filled with cold dust, possibly formed after the passage of the shock front.

%Section 5
\section{Discussion}
\label{discussion}

%Figure 2
\begin{figure}
\includegraphics[width=0.475\textwidth]{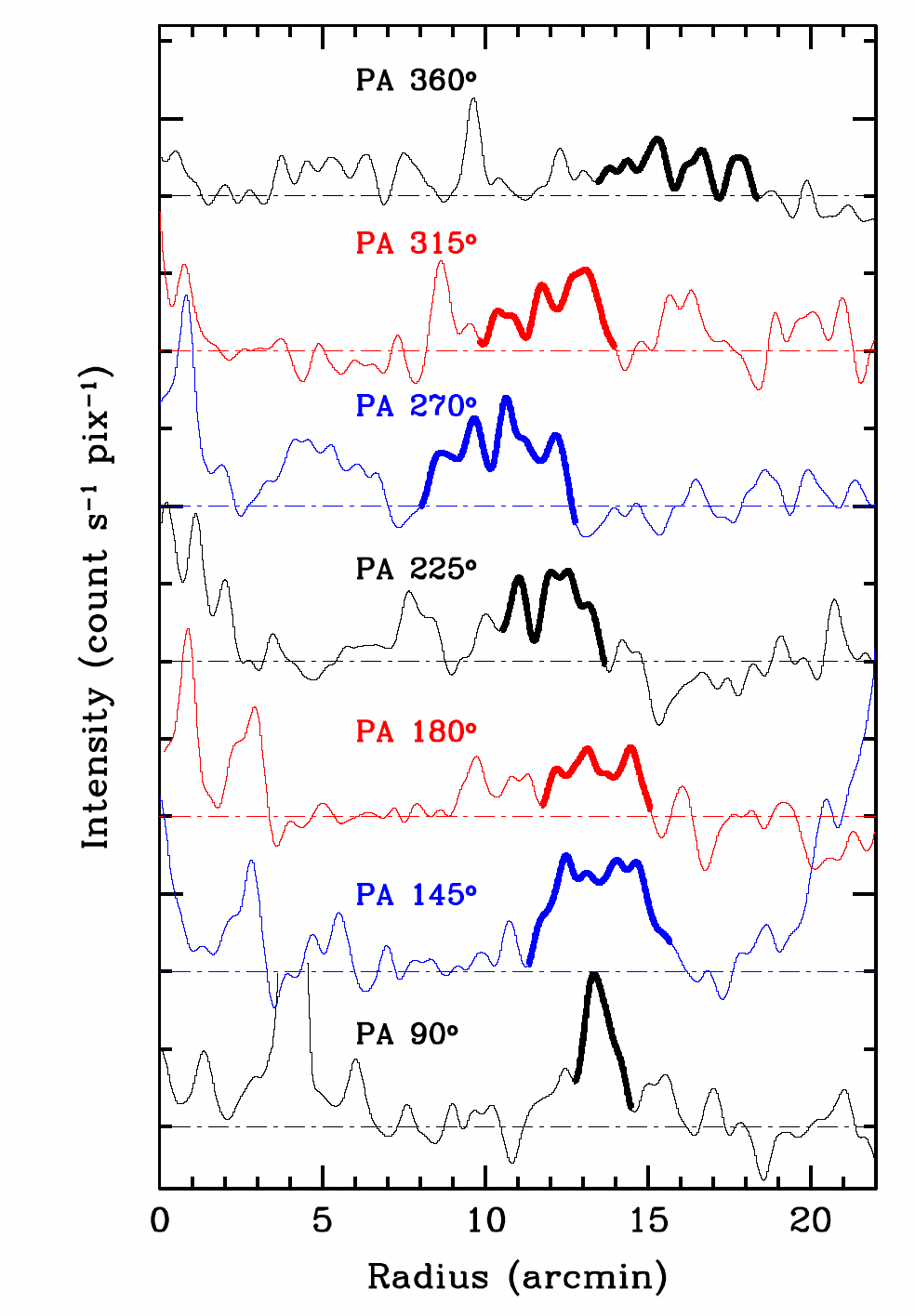}
\caption{
{\it GALEX} far-UV surface brightness profiles of R\,Dor along a number of radial directions.  
The profile along PA = 45$^\circ$ is not shown due to the strong contamination of a bright star.  
Similarly, the profile along PA = 135$^\circ$ has been tweaked to 145$^\circ$ to avoid a background star.  
The vertical tickmarks correspond to 5$\times$10$^{-7}$ cnt~s$^{-1}$pix$^{-1}$ and each profile is shifted vertically by 1$\times$10$^{-6}$ cnt~s$^{-1}$pix$^{-1}$ with respect to the previous one.  
% The vertical solid line marks the average radius of the shell and the 
The dot-dashed horizontal lines mark the background emission level. 
The extent of the shell at each profile is highlighted using thick lines. 
}
\label{profile}
\end{figure}

%Section 5.1
\subsection{On the detection of UV bubbles around AGB stars}
\label{conditions}

The presence of shells around mass-losing stars has long been associated with episodes of increased mass-loss. 
Actually \citet{Nhung2019} argue that R\,Dor experienced a recent episode of intense mass loss about $10^2$ yr ago, as they found evidence for strong inhomogeneities in the distribution of SO between 20 and 100 AU from the star. Since episodes of enhanced mass loss happen customarily during the AGB phase, circumstellar structures can be found around numerous AGB stars. These are detected mostly in far-IR observations \citep{Cox2012}, whereas the number of those detected in UV observations is conspicuously small. In this section we will analyse the observational limitations for their UV detection.

%Figure 3
\begin{figure}
\includegraphics[width=0.475\textwidth]{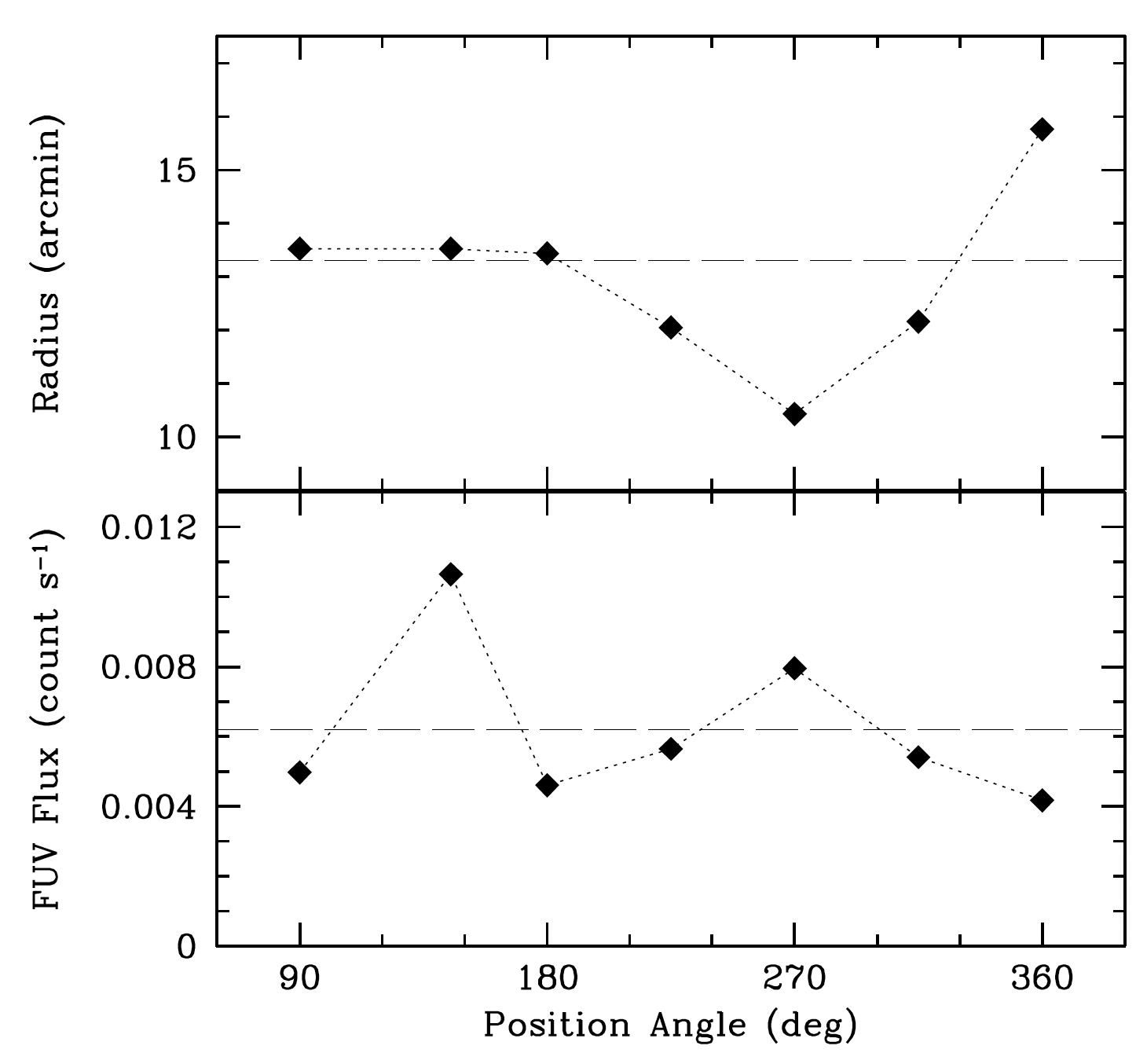}
\caption{
Azimuthal dependence of the flux (bottom) and radius (top) of the {\it GALEX} far-UV emission of R\,Dor derived from Figure~\ref{profile}. The flux is integrated from a 3$^\prime$-wide aperture. The horizontal dashed lines mark the average flux (bottom panel) and radius of the shell (top panel). 
}
\label{radial}
\end{figure}

Since the first detection of a UV detached shell in the form of a cometary tail around $o$ Ceti \citep{Martin2007}, it became evident that the detection is better achieved using far-UV images. 
Apart the possibility that the bubbles can be intrinsically brighter in the far-UV, the interstellar extinction at this wavelength is actually {\it lower} than in the near-UV. This happens because the \emph{GALEX} near-UV band is centred at a wavelength very close to the ``UV bump'' at 2175 \AA \, \citep{Cardelli1989}.
For example, the bubble around CIT\,6 \citep{SM2014} appears very sharp in the far-UV, but it is absent in the near-UV {\it GALEX} images, alike R Dor. 
\citet{SC2010} estimated the far-to-near-UV brightness ratio $I_{\rm FUV}/I_{\rm NUV} \simeq 6$ at the rim of the bubble associated with IRC+10216. Assuming that this same ratio is valid for R Dor the near-UV peak intensity at the rim of this bubble should be $I_{\rm NUV} \la 2 \times 10^{-7}$ count s$^{-1}$ pix$^{-1}$, which is within the noise level of the near-UV image (Fig. \ref{profile}).
Unfortunately, the vast majority of {\it GALEX} images are in the near-UV, where the bubbles appear fainter. 
As a result, only a minor fraction of the AGB stars observed by {\it GALEX} are candidates for new detections. 
Increasing the number of far-UV observations of AGB stars could possibly lead to an increase in the number of detections.

\begin{figure*}
\includegraphics[width=0.95\textwidth]{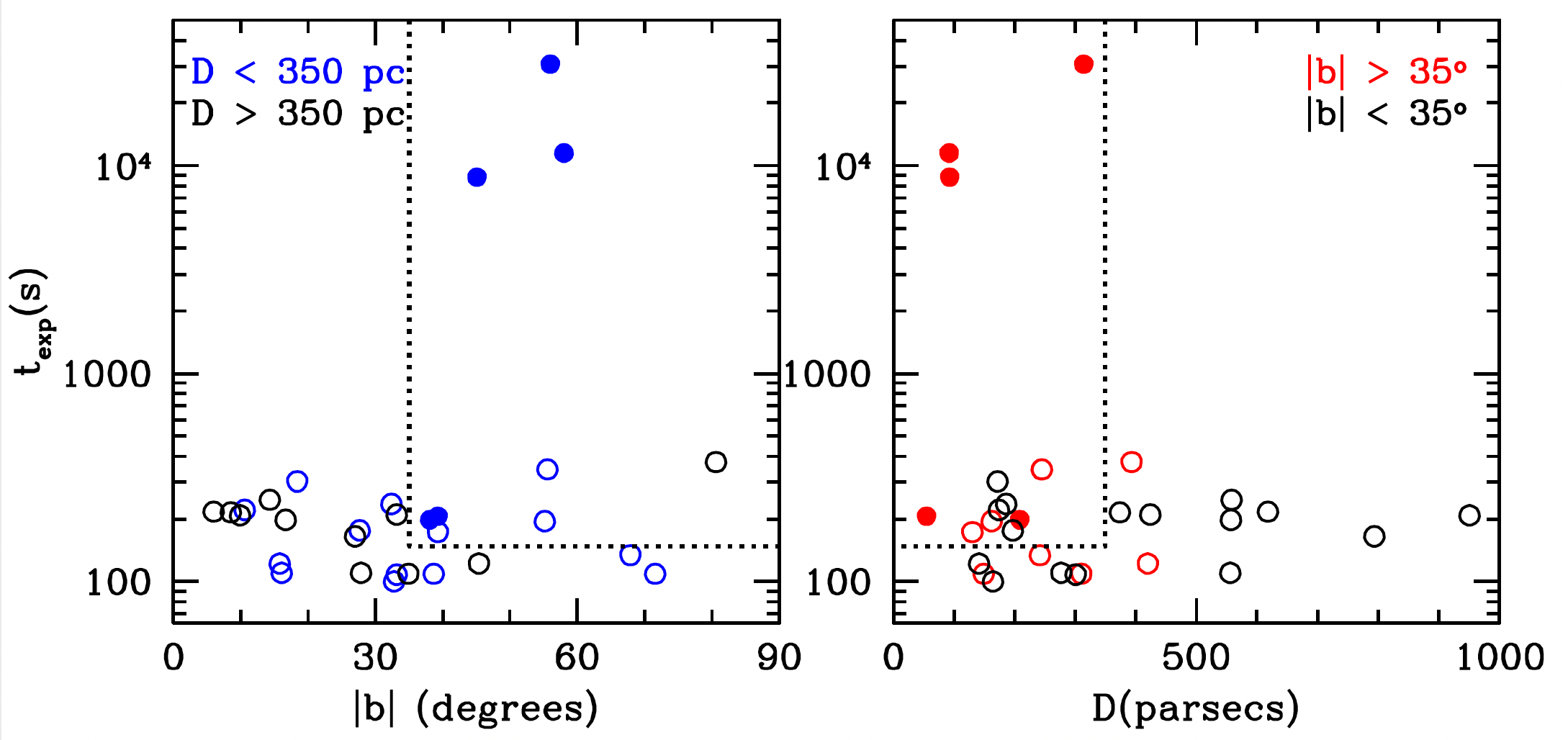}
\caption{
Galactic latitude $b$ (left) and distance $D$ (right) distributions of the AGB stars with large-scale IR \citep{Cox2012} and UV (this paper and references in Tab.\ref{listado}) shells versus their \emph{GALEX} far-UV exposure times ($t_{\rm exp}$). 
Empty symbols correspond to sources with IR only shells and filled symbols to UV shells.  
The dotted lines mark the loci where UV bubbles are found: $D \la 350$ parsec, $|b| \ga 35\degr$, and $t_{\rm exp} \ga 150$ s. 
In the left panel, shells within 350 parsec are marked in blue, whereas in the right panel shells with $|b| > 35\degr$ are marked in red.  
Three IR only shells, namely EP Aqr, TX Psc and $\pi^1$ Gru, are within 350 parsec and closer than $35\degr$ to the Galactic plane, yet they are not detected in the available \emph{GALEX} far-UV images.  
}
\label{cox2}
\end{figure*}

Secondly, in order to have its detached shell detected in the UV, the AGB star must be relatively near. 
Newly born bubbles (i.e. recently ejected by the star), which are expected to be brighter, require high resolution images to be detected, especially if the star is far (and vice-versa). 
Moreover UV radiation is severely affected by interstellar extinction, which introduces a strong bias towards the detection of nearby objects. 
Table~\ref{listado} and Figure~\ref{cox2} confirm that $o$ Cet and the four bubbles detected so far are all close, within a distance $\la 350$ parsec, and have high galactic latitude ($\vert b \vert > 35\degr$). 
The farthest object in the list, CIT 6, was not detected in the near-UV, perhaps because of the relatively higher extinction towards it. 
Moderate-to-high interstellar extinction, which increases with distance and along the Galactic Plane, hinders the detection of UV bubbles around AGB stars.
% Since these bubbles appear less evident in the near-UV images moderate-to-high interstellar extinction can make their detection more difficult.

The third reason that possibly plays a role in the detection of these UV bubbles is the sensitivity of the observations. Presently, the {\it GALEX} images constitute the main database available for this purpose. 
However, the exposure time ($t_{\rm exp}$) of the vast majority of these observations is only $\sim 10^2$ s long. 
Until the discovery of the UV bubble around U Hya bubble, detected in a shallow {\it GALEX} far-UV image, the previous detections of IRC+10216 and CIT 6 were based on observations at least 100 times longer (Tab.~\ref{listado}). 
% Until the discovery of the bubble around CIT 6 in 2014, long exposures were considered mandatory to detect these bubbles. 
% Eventually, the U Hya bubble was detected in a shallow {\it GALEX} far-UV image with $t_{\rm FUV}$ two orders of magnitude shorter than the previous findings, IRC+10216 and CIT 6 (Table \ref{listado}). 
The compact morphology of U Hya might have contributed to its detection, 
% because the more compact is the shell, the brighter. 
but the present discovery of the largest angular diameter (26\farcm6) bubble ever found in a short (only 206 s) {\it GALEX} far-UV image casts doubts on the importance of the role of the exposure time on the detection of bubbles. 
% It would seem that shallow {\it GALEX AIS} images can also be efficient to detect these bubbles, even though some of the fainter bubbles might remain overlooked. \\
It is worth investigating whether shallow {\it GALEX AIS} images can also be efficient to detect these bubbles.  

Figure \ref{cox2} represents graphically these observational conditions described above. 
The plot shows \emph{GALEX} far-UV exposure times, galactic latitudes, and distances of the UV bubbles of Table \ref{listado} and the IR shells by \citet{Cox2012}. 
A dotted line marks the region in the diagrams where UV bubbles have been found, i.e. $t_{\rm exp} \ga 150$ s, $b \ga 35\degr$ and/or $D \la 350$ parsecs. Only three IR shells without a UV counterpart are found inside the loci of the UV shells: EP Aqr, TX Psc, and $\pi^1$ Gru. Apart from these few exceptions, all of them with small angular diameters ($\la 1\arcmin$), our criteria can explain the detections of UV bubbles among sources with IR shells.  

%Figure 4

In short, discovering a circumstellar bubble around an AGB star depends on two factors: (1) the observational conditions for its detection, as discussed above; and (2) the actual existence of the bubble, which in turn depends on the physical processes that contribute to its formation and endurance. In Sect. \ref{processes} we briefly discuss some of these phenomena.

\begin{table*}
 \caption{Main characteristics of UV detached shells around AGB stars known to date, in order of discovery. Except $o$ Cet, all shells are approximately round and are denominated in this work ``bubbles''. $H$ is the distance from the Galactic plane; $<\phi>$ and $<$Diam.$>$ are the apparent and actual diameters, respectively; $V_t$ and $V_r$ are the tangential and heliocentric radial velocities, respectively, extracted from the {\sc simbad} database \citep[except IRC+10216,][]{Menzies2006}; $V_{UVW} = \sqrt{ U^2+V^2+W^2}$ is the stellar galactic space velocity relative to the local standard of rest (not corrected for the solar motion); $t_{\rm exp}$ is the exposure time of the {\it GALEX} far-UV images. 
 References in the last column are: M2007 = \citet{Martin2007}, SC2010 = \citet{SC2010}, SM2014 = \citet{SM2014}, S2015 = \citet{Sanchez2015}, OG2023 = this work.}
  \begin{tabular}{lrrcccccccl}
 \hline
Name & 
\multicolumn{1}{c}{Distance} & 
\multicolumn{1}{c}{$H$} & 
\multicolumn{1}{c}{$<\phi>$} & 
\multicolumn{1}{c}{$<$Diam.$>$} & 
$b$ & $V_t$ & $V_r$ & $V_{UVW}$ & $t_{\rm exp}$ & Reference \\
 & 
\multicolumn{1}{c}{(parsec)} & 
\multicolumn{1}{c}{(parsec)} & 
\multicolumn{1}{c}{(arcmin)} & 
\multicolumn{1}{c}{(parsec)} & 
(deg) & (km s$^{-1}$) & (km s$^{-1}$) & (km s$^{-1}$) & (s) & \\
\hline
$o$ Cet & 91.7~~ & 77.8~~ & - & - & $-58.0$ & 103.2 & $+63.5$ & 121.2 & $11.5 \times 10^3$ & M2007 \\
IRC+10216 & 92.7~~ & 65.7~~ & $24\arcmin$ & 0.65 & $+45.1$ & 15.5 & $-23.2^a$ & 27.9 & $8.8 \times 10^3$ & SC2010 \\
CIT 6 & 314.1~~ & 260.4~~ & $15\arcmin \times 18\arcmin$& $1.37 \times 1.64$ & $+56.0$ & 31.0 & $-0.3$ & 31.0 & $30.8 \times 10^3$ & SM2014 \\
U Hya & 208.3~~ & 128.5~~ & $3.7\arcmin$ & 0.22 & $+38.1$ & 56.2 & $-25.8$ & 61.8 & 198 & S2015 \\
R Dor & 54.6~~ & 34.6~~ & $26\arcmin \times 29\arcmin$ & $0.41 \times 0.46$ & $-39.3$ & 26.6 & $+26.1$ & 37.2 & 206 & OG2023 \\
\hline
\end{tabular}
\label{listado}
\end{table*}

%Section 5.2
\subsection{On the formation of UV bubbles around AGB stars}
\label{processes}

\citet{Cox2012} report the detection at 70 and 160 $\mu$m of numerous bow shocks and detached shells caused by the interaction between the stellar wind of AGB stars and supergiants and the ISM.  
They also predicted that these structures might be eventually found towards another eight stars within the distance of 500 parsecs, namely TW Hor, V Eri, R Dor (this work), R Lep, RS Cnc, HD100764, RY Dra, and RX Boo.

Bubbles are formed from the shock between the stellar wind and the ISM \citep{SC2010}.  
The brightness of the shocked gas generally depends, among various factors, on the product of the density of the ISM and the stellar wind. 
Except for CIT 6, all the other AGB stars with UV bubbles are within 150 parsecs from the Galactic plane, which is the H~{\sc i} scale height \citep{KK2009}. 
Therefore, the \ion{H}{i} density near those bubbles is in the range between 0.1 and 1 cm$^{-3}$, which is a narrower interval (1 dex) than that of mass loss rate values during the AGB phase \citep[3 dex, from $10^{-7} \sim 10^{-4} M_{\odot}$~yr$^{-1}$,][]{HO2018}. 
We conclude that the higher is the mass loss rate, the brighter is the bubble. 
IRC+10216 is a nearby AGB star showing a very high mass loss rate \citep[$\dot{M}=2 \times 10^{-5} M_{\odot}$~yr$^{-1}$,][]{CM1997,G1998}, whilst the mass loss rate of R Dor is two orders of magnitude lower ($(0.6 - 1.6) \times 10^{-7} M_{\odot}$~yr$^{-1}$, see Sect.~\ref{circumstellar} for references) and that of U Hya is even lower \citep[$\dot{M}=4.9 \times 10^{-8}M_{\odot}$yr$^{-1}$,][]{DeBeck2010}. 
These values, however, refer to the ``quiet'' mass loss phase, i.e. a period of time when the star loses mass at a constant rate, which covers over 90\% of the time during the AGB phase. Just after a thermal pulse, the AGB star may undergo an episode of intense mass loss, forming an expanding shell of dense gas. 
Therefore, the formation of a bubble would depend mainly on the density of the wind {\it during the episode of enhanced mass loss}, and secondly on the density of the local ISM. 
This can explain why both AGB stars with relatively low (R Dor, U\,Hya) and high (IRC+10216, CIT 6) mass loss rates during the quiet phase may exhibit a bubble.

In addition to thermal pulses, with typical time lapse of $\sim 10^5$ yr between two consecutive thermal pulses \citep{VW1993}, {\it brief episodes of enhanced mass loss} may occur at time scales of a few hundred years \citep{HO2018}. CIT 6 is the largest bubble known to date ($<$Diam$>=1.37 \times 1.64$ parsec$^2$), expanding at a speed of 18 km s$^{-1}$ \citep{SM2014}. At this speed, the kinematical age (size divided by velocity) of this bubble implies that it was ejected $\sim 8.2 \times 10^4$ yr ago. The same calculus applied to R Dor, assuming its present-day wind speed of 6.0 km s$^{-1}$, results in $3.4 \times 10^4$ yr. These ages are within the same order of magnitude than the time lapse of $\sim 10^5$ yr between consecutive thermal pulses, although we note that the present-day measurements of the stellar wind velocity of a few km s$^{-1}$ during the ``quiet phase'' (i.e. between the episodes) might not necessarily reflect the enhanced velocity of the stellar wind during those episodes \citep{VW1993}. Since the formation of shells associated with episodes of enhanced mass loss is expected to be common among AGB stars, the small number of UV bubbles known to date shall be attributed to the observational biases described in Sect. \ref{conditions}. 

%Figure 5

%Section 5.3
\subsection{The origin of the far-UV emission}

When studying the shell around U Hya, \citet{Sanchez2015} argued that scattering of UV photons originated in the central star or in the interstellar radiation field does not account for the necessary photons corresponding to the UV luminosity of the bubble associated to U Hya. 
Therefore, the excitation of atoms or molecules by the shock between the stellar wind and the local ISM remains the most likely mechanism to explain the UV emission of the shell \citep{SC2010}.

Some AGB stars with extended shells exhibit significant velocities relative to the local medium. Let us assume that the ISM is not moving relative to the local standard of rest, then the galactic space velocity can be derived as $V_{UVW}=\sqrt{U^2+V^2+W^2}$. 
Table \ref{listado} shows that, except the fast-moving star $o$ Cet, the remaining AGB stars with UV bubbles all move at $V_{UVW} = 28 \sim 62$ km s$^{-1}$.
These values are almost one order of magnitude higher than the typical velocity of the stellar wind of AGB stars.  \citet{Sanchez2015} argue that the post-shock temperature can reach $T_{\rm ps} \simeq 1.6 \times 10^5$ K $({\mu}_{\rm H}/1.33)(V_*/72$ km s$^{-1})$, for a mean molecular mass ${\mu}_{\rm H}$. 
This suggests that the relatively high velocity of the star (and its wind) relative to the ISM could be responsible for the raise of temperature necessary to excite the atomic species, and produce the emission lines in the UV.

Unfortunately, no UV spectra of extended shells around AGB stars have been obtained to date, thus the spectral features associated with the UV emission cannot be unequivocally identified. 
The {\it GALEX} far-UV filter extends from 1344 to 1786 \AA, which does not include  Lyman-$\alpha$, the strongest emission line in the UV domain. 
Nevertheless, there are plenty of other species showing emission lines in this spectral interval, such as: (1) low-excitation CO emission lines; 
% , like those observed in the chromospheric far-UV spectrum of red giants \citep{Carpenter1988}; 
(2) low-ionization ions, formed at temperatures of $T = 10,000 \sim 15,000$ K (e.g. various \ion{Fe}{ii} lines, \ion{O}{i}$ \lambda \lambda 1355,1640$, \ion{Si}{ii} $\lambda 1529$, \ion{C}{i}$ \lambda 1657$, \ion{O}{iii} $\lambda 1663$); (3) a few higher ionization species, which require higher temperatures ($T = 35,000 \sim 50,000$) like the doublets \ion{Si}{iv}$ \lambda 1400$ and \ion{C}{iv}$ \lambda 1550$, for example. Most of these features have been identified in the chromospheres of giant late-type stars \citep{JJ1984,Carpenter1988,Ortiz2019}, which show temperatures of the same order of magnitude as the post-shock temperatures estimated above.

%Section 5.4
\subsection{The mass of the UV thin bubble}

The mass of the thin bubble can be roughly estimated assuming that the thin shell seen in the UV corresponds to a zone where a shock wave has been formed after an episode of enhanced mass loss. 
The mass loss during the quiet phase is given by the following equation:

\begin{equation}
\dot{M}=4 \pi r^2{\rho}_w v_w,
\label{eq1}
\end{equation}

\noindent 
where ${\rho}_w$ and $v_w$ are the mass density and velocity of the stellar wind, respectively. For an adiabatic shock of a monoatomic gas (i.e. $\gamma = 5/3$), the shock/wind density ratio is:

\begin{equation}
    \frac{{\rho}_s}{{\rho}_w}=\frac{\gamma +1}{\gamma -1}=4.
\label{gamma}
\end{equation}

\noindent 
From these two previous equations we get:

\begin{equation}
    {\rho}_w=\frac{\dot{M}}{4 \pi r^2 v_w}=\frac{1}{4}{{\rho}_s}.
\end{equation}

\noindent Let us assume the UV thin shell has thickness $l$ at radius $r$. Its mass can be estimated as follows:

\begin{equation}
    M_s \simeq 4 \pi r^2 l {\rho}_s = \frac{4 \dot{M} l}{v_w}.
\end{equation}

\noindent Assuming $\dot{M}\simeq 10^{-7} M_{\odot}$yr$^{-1}$, $v_w=6$ km s$^{-1}$, and the angular thickness of the bubble of $\sim $3\arcmin \, (from Fig. \ref{profile}) at a distance of 54.6 parsecs, we obtain $M_s=0.0031 M_{\odot}$. Therefore, if bubbles like this are formed because of the thermal pulses, after the ten or twenty episodes expected to happen during the AGB phase the total mass lost due to this phenomena must not exceed $\approx 0.06 M_{\odot}$. Although this mass loss may have consequences for the stellar evolution, it is over one order of magnitude less than the total mass loss during the ``quiet'' (i.e. non-episodic) AGB phase.

\section{Conclusions}
\label{conclusions}

In this paper we describe the discovery of a bubble around the AGB star R Dor detected in the \emph{GALEX} far-UV band. This is the fourth detection of a UV bubble around an AGB star after IRC10216, CIT6, and U Hya ($o$ Ceti has a ``cometary tail'', a morphologically different envelope). 
Apart from $o$ Ceti, R Dor is the only O-rich star among these to show a UV extended feature. 
These are the main characteristics of the bubble:

\begin{itemize}

\item 
Its shape is approximately round, with an actual diameter of $0.41 \times 0.46$ parsec$^2$. The angle between the minor axis of this ellipse and the stellar proper motion is $\sim 28\degr$. Thus, we cannot establish a firm relationship between the shape of the bubble and the movement of the central star relative to the ISM.

\item
Assuming that the UV emitting region corresponds to a shock zone caused by an episodic mass loss episode, its mass has been estimated as 0.0031 $M_{\odot}$. Considering that an AGB star can undergo between ten to twenty thermal pulses, the total mass loss due to this phenomena is less than $\approx 0.06 M_{\odot}$. This is over one order of magnitude less than the total mass loss during the ``quiet'' (i.e. between pulses) AGB phase.

\item 
The bubble, like other cases previously reported in the literature, is filled with far-IR radiation.  
The 60 and 100 $\mu$m \emph{IRAS} fluxes are consistent with a colour temperature $\la 32$K. 
This is compatible with the presence of cold dust, probably formed after the passage of the shock front. 

\end{itemize}

Large scale bubbles around AGB stars are the consequence of episodes of intense mass loss.  
The kinematical ages of these bubbles are consistent with the time lapse between thermal pulses.  
Accordingly, a significant fraction of the AGB stars can be expected to exhibit a UV bubble. 
The small number of cases reported in the literature would be the result of observational biases. 
% that make difficult the detection of UV bubbles beyond a few hundred parsecs of distance.
Since all UV bubbles are found to be within 350 parsec and at high galactic latitudes ($|b| \ga 35\degr$), the interstellar absorption in the UV is considered to be the main cause preventing the discovery of more UV bubbles around AGB stars. Deeper far-UV observations can reveal those. 

\section*{Acknowledgements}

We acknowledge with thanks the variable star observations from the {\it AAVSO International Database} contributed by observers worldwide and used in this research.
M.A.G.\ acknowledges financial support from grants CEX2021-001131-S funded by MCIN/AEI/10.13039/501100011033 and PGC2018-102184-B-100 from the Spanish Ministerio de Ciencia, Innovaci\'on y Universidades (MCIU). 
This research has made extensive use of the SIMBAD database, operated at CDS, Strasbourg, France, and NASA's Astrophysics Data System. We thank the anonymous referee for his/her helpful comments and suggestions.

\section{Data availability}

The data underlying this article are publicy available in: {\it The IRAS Sky Survey Atlas (ISSA)}, hosted at the IRSA website {\it https://irsa.ipac.caltech.edu/Missions/iras.html}; the {\it GALEX tile search}, hosted by {\it The Barbara Mizulski Archive for Space Telescopes}, at the {\it GALEX} website {\it http://galex.stsci.edu/gr6/?page=tilelist\&survey=allsurveys}. Visual photometric data archived by the {\it American Association of Variable Stars Observers, AAVSO} are available at their website {\it www.aavso.org}.

% Don't change these lines
\bsp    % typesetting comment
\label{lastpage}
\end{document}